\documentclass[a4paper,11pt]{article}
\pdfoutput=1 

\usepackage{jheppub} 

\usepackage[T1]{fontenc} 

\title{\boldmath NLO QCD corrections to Higgs boson production in association with gauge bosons in proton-proton collisions at $\sqrt{s}$ = 14 TeV.}

\author[a]{K.Djamaa,}
\author[a]{A. Mohamed-Meziani}

\affiliation[a]{Department of Physics, Faculty of Exact Sciences, University of Bejaia,06000 Bejaia, Algeria}

\emailAdd{kenza.djamaa@univ-bejaia.dz}
\emailAdd{abdelkader.mohamedmeziani@univ-bejaia.dz}

\abstract{  We suggest some predictions for Standard Model Higgs boson production in association with gauge bosons in proton-proton collisions at $\sqrt{s}$ = 14 TeV.  Our calculation includes the NLO QCD corrections to these processes using MadGraph5$\_$aMC@NLO event generator. We study the impacts of these corrections on the total cross sections and various  kinematical distributions.
 }

\begin{document} 
\maketitle
\flushbottom

\section{Introduction }
\label{sec:1}
The discovery of Higgs boson at LHC on 2012~\cite{ref:1,ref:2} is the key endeavour for various phenomena in high energy physics. Hence, it is the source of electroweak symmetry breaking mechanism and the mass generation of the Standard Model (SM). The production of the Higgs boson associated with gauge bosons W$^{\pm}$ or Z( also called Higgs-strahlung process) is one of the important processes of the Higgs production modes, it provides informations about the the Higgs couplings with the vector boson and other particles of the SM and it can be used to search signals for Beyond Standard Model  physics. For these reasons, the processes pp $\rightarrow$ VH (V= W$^{\pm}$/ Z ) have a significant interest at the ATLAS and CMS collaborations. The results of VH production with the ATLAS detector were performed in~\cite{ref:3} at centre-of-mass energies $\sqrt{s}$ = 7 TeV and 8 TeV. At $\sqrt{s}$ = 13 TeV the ATLAS collaboration has analysed  data collected between 2015 and 2016~\cite{ref:4}, 2015 and 2018~\cite{ref:5}. The CMS experiment has presented a search for study the VH production at $\sqrt{s}$ = 8 TeV and 13 TeV corresponding to respective integrated luminosities of 19.7 and up to 2.7 fb$^{-1}$~\cite{ref:6}.

In last past years, a large number of theoretical predictions for H production associated to gauge bosons have been performed, firstly at leading order only via $q\bar{q}$~\cite{ref:7,ref:8}, then at next-to-leading order with QCD and QED corrections  have been presented in ~\cite{ref:9,ref:10} and~\cite{ref:11} respectively. Computations of  next-to-next-to-leading order QCD corrections to  pp $\rightarrow$ VH have been implemented in~\cite{ref:12}. The off-shell of NLO for WH production (  H$\rightarrow$ $b\bar{b}$, W $\rightarrow$ $l\nu_{l}$) has been studied in~\cite{ref:13} at $ \sqrt{s}$ = 8 TeV and 14 TeV using MC@NLO simulation. The association of a jet to the WH production was evaluated in~\cite{ref:14} at$ \sqrt{s}$ = 13 TeV within
the NNLOJET framework and  NNPDF31 from the LHAPDF library. Much attention has given to ZH production which was computed in~\cite{ref:15} at NNLO QCD corrections employing the vh@nnlo program at $ \sqrt{s}$ = 8 TeV and 14 TeV. ~\cite{ref:16} discusses the processes of ZH production through gluon-gluon and photon-photon collisions.

This paper focuses on the VH production in proton-proton collisions at $ \sqrt{s}$ = 14 TeV. We have carried out these processes at LO using MadGraph5$\_$aMC@NLO~\cite{ref:17} event generator, then we have included the NLO QCD corrections to exploit these processes at high precision. We also report numerical predictions of their LO and NLO total cross sections and the corresponding scale uncertainty. Furthermore, we present the several distributions in  transverse momentum of gauge vectors, two tagging jets and missing energy transverse after showering and hadronisation using Pythia8~\cite{ref:18}.

The remainder of this article is organized as follows. After this brief introduction, we discuss the effect of the NLO QCD corrections on the contributions of VH production  in section~\ref{sec:2}. In section~\ref{sec:3}, we present our phenomenological results on the total cross section and various distributions. Finally  we  give our conclusions in section~\ref{sec:4}.

\section{ Details of the calculation}
\label{sec:2}

In this section, we focus on the production on-shell of the processes 
\begin{eqnarray}
\label{eq:1}
 p p \rightarrow VH 
\end{eqnarray}

where V= W$^{\pm}$, Z. We work in the Standard Model with 4-flavor scheme, i.e. p = u, d, c, s and gluons. We take into account the virtual bottom-loop, top-loop but not their  contributions in the initial state. Feynman diagrams are generated at LO and NLO with QCD corrections using MadGraph5$\_$aMC@NLO version 2.6.0.

\begin{figure}
	\centering	
	\includegraphics[width=.30\textwidth,trim=0 0 0 0,clip]{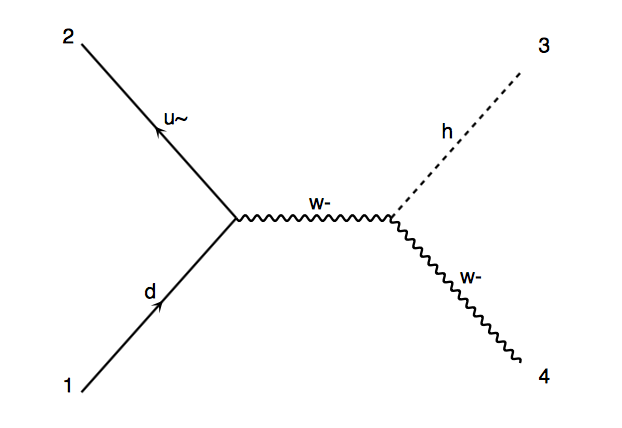}	\includegraphics[width=.30\textwidth,trim=0 0 0 0,clip]{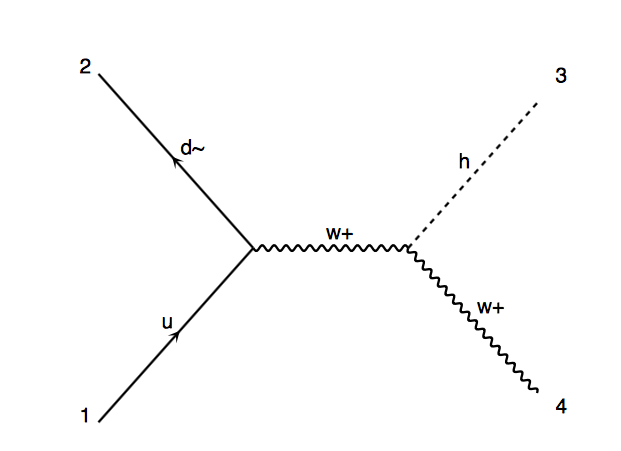} 	
	\includegraphics[width=.30\textwidth,trim=0 0 0 0,clip]{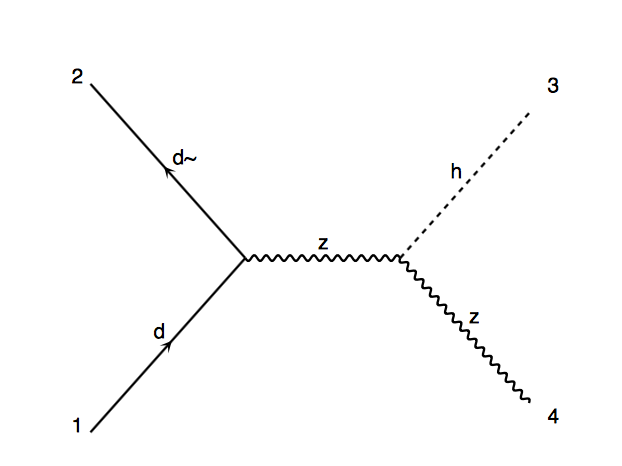} \\
	a \qquad \qquad \qquad \qquad  \qquad \qquad b  \qquad \qquad \qquad \qquad  \qquad \qquad c \qquad \qquad \qquad \qquad

	\caption{Examples of Feynman diagrams for the different processes of VH production at tree-level.\label{fig:1}
	}	
\end{figure}

We start by generating the processes at leading order, where the process pp $\rightarrow$ W$^{\pm}$H is exclusively via by the $t$-channel of $q\bar{q'}$ annihilation with exchange of W$^{\pm}$, as shown in figures~\ref{fig:1}a and b.  However, the $q\bar{q}$ is the only annihilation that contributes in the production of ZH and Z  is the exchanged particle as displayed in figure~\ref{fig:1} c.  

The $O(\alpha_{s})$ NLO QCD corrections consist new channels with a real and virtual contributions. They involve infrared (IR) and ultraviolet (UV) divergences which are regulated them using the dimensional regularization (DR) scheme ~\cite{ref:19}. 

\begin{figure}
	\centering	
	\includegraphics[width=.30\textwidth,trim=0 0 0 0,clip]{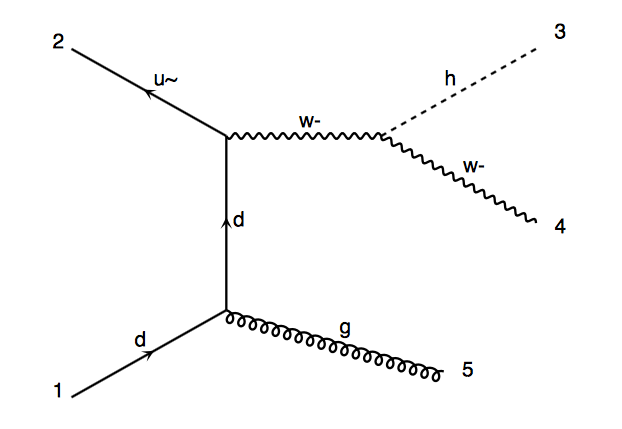}	\includegraphics[width=.30\textwidth,trim=0 0 0 0,clip]{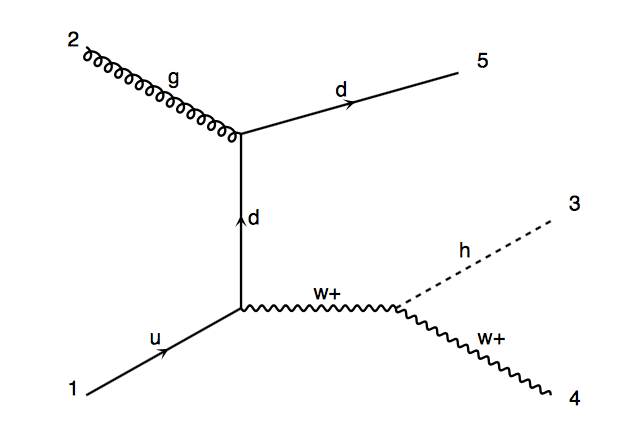} 	
	\includegraphics[width=.30\textwidth,trim=0 0 0 0,clip]{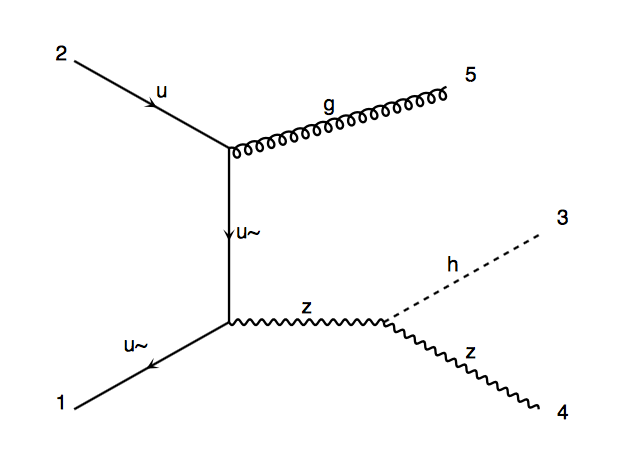} \\
	a \qquad \qquad \qquad \qquad  \qquad \qquad b  \qquad \qquad \qquad \qquad  \qquad \qquad c \qquad \qquad \qquad \qquad

	\caption{Examples of Feynman diagrams of real corrections for the different processes of VH production.\label{fig:2}
	}	
\end{figure}

The real emission for VH production is essentially the tree-level diagrams with an additional parton and we classify it into two groups: $q\bar{q'} \rightarrow $ W$^{\pm}$H $g$ and $q(\bar{q'})g \rightarrow $ W$^{\pm}$H $q(\bar{q'})$  some examples are given in figures~\ref{fig:2} a and b. The same is true for pp $\rightarrow$ ZH except that $q' = q$ as represented in figure~\ref{fig:2} c. 

\begin{figure}
	\centering	
	\includegraphics[width=.30\textwidth,trim=0 0 0 0,clip]{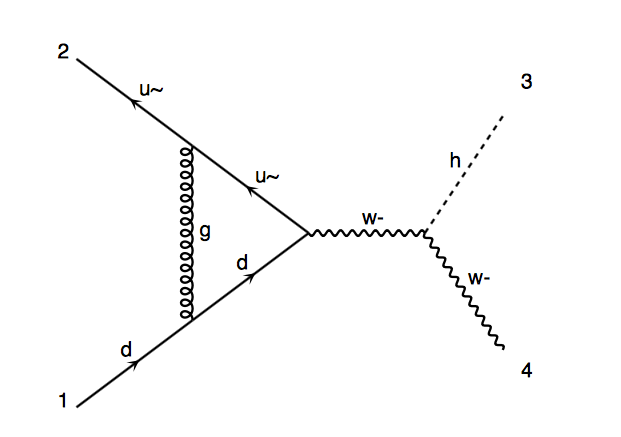}	\includegraphics[width=.30\textwidth,trim=0 0 0 0,clip]{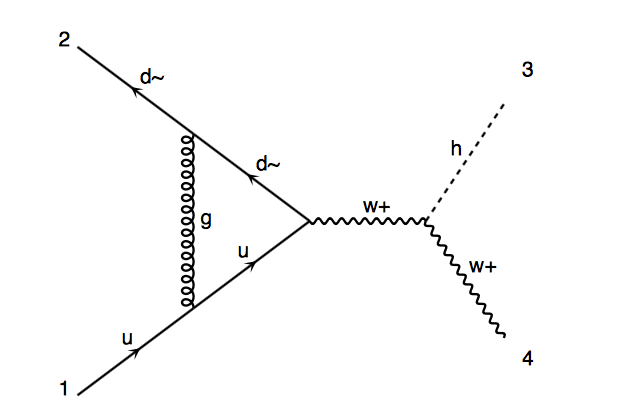} 	
	\includegraphics[width=.30\textwidth,trim=0 0 0 0,clip]{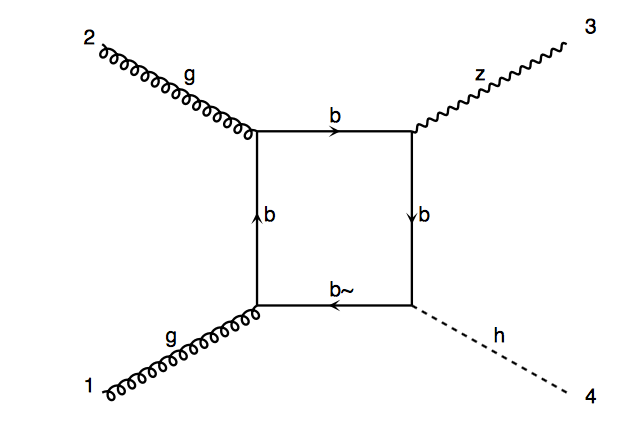} \\
	a \qquad \qquad \qquad \qquad  \qquad \qquad b  \qquad \qquad \qquad \qquad  \qquad \qquad c \qquad \qquad \qquad \qquad

	\caption{Examples of Feynman diagrams of virtual corrections for the different processes of VH production. \label{fig:3}
	}	
\end{figure}

The virtual corrections are also computed, we have noticed  that the one-loop of pp $\rightarrow$ W$^{\pm}$ H has triangle form and it is only obtained by $q\bar{q'}$ initiated as illustrated in figures~\ref{fig:3} a and b. While, the ZH production has two types of loops boxes, triangles and they are obtained by $q\bar{q}$  and gg contributions as shown in ~\ref{fig:3} c. The gluon-initiated produces a top-quark and bottom loops which represent an interesting feature of dominant contributions in the SM and beyond the SM. 

\section{Phenomenological results  and discussion }
\label{sec:3}

In the following we present predictions of the total cross section and the distributions of kinematics variables for VH production at LHC in the case of $\sqrt{s} $= 14 TeV, using PYTHIA8 to showering and hadronization the events. We employ NNPDF23 set parton distribution functions from LHAPDF setup ~\cite{ref:20}, adopting the strong coupling constant $\alpha_{s}$($M_{Z} $) = 0.1137 and 0.1251 at LO and NLO respectively. Moreover, we take the central numerical value of the renormalization and factorization scales set to $\mu_{R}$ = $\mu_{F}$ = $M_{H} $.

we collect the SM input parameters using for all numerical results:

$G_{\mu}$ = 1.16637.$10^{-5}$ GeV$^{-2}$,  \qquad \qquad $\alpha^{-1}$ = 132.338; 

$M_{W} $ = 80.385 GeV,  \quad \qquad \qquad \qquad $\Gamma_{W}$ = 2.085 GeV;

$M_{Z} $ = 91.188 GeV,  \quad \qquad  \qquad \qquad $\Gamma_{Z}$ = 2.495 GeV; 

$M_{H} $ = 125 GeV,   \qquad \qquad \qquad \qquad $\Gamma_{H}$ = 0.00407 GeV; 
 
$M_{t} $ = 173.2 GeV,   \qquad \qquad  \qquad \qquad $\Gamma_{t}$ = 1.4426 GeV; 

Meanwhile, we impose the following cuts on transverse momentum, pseudo-rapidity of the leptons and jets:
\begin{equation}
\label{eq:2}
\ |\eta(l)| \leq 2.5, \quad  p_{ T}(l) > 20  GeV \nonumber
\end{equation}
\begin{equation}
\label{eq:3}
\ |\eta(j)| \leq 2.5, \quad  p_{ T}(j) > 10  GeV
\end{equation}
Jets are reconstructed with the anti-kT algorithm~\cite{ref:21} with a radius parameter set to R = 0.6.

\subsection{Total cross sections}

\label{subsec:3.1}

\begin{table}
	\centering
	\begin{tabular}{|c|c|c|c|}
		\hline

		&	$ \sigma_{LO}$[pb]  & $\sigma_{NLO}$[pb]  & $ K =  \frac{\sigma_{NLO}}{\sigma_{LO}}$      \\ \hline 
		
		W$^{-}$H     &  0.44$ \pm5.2\times10^{-4}$ $^{+3.58\%} _{-4.37\%}$ & 0.605$ \pm 1.5\times10^{-3}$ $^{+1.7\%} _{-2.3\%}$         &  1.38   \\  \hline
		
		W$^{+}$H  &  0.717 $ \pm 5.3\times 10^{-4}$  $^{+3.22\%} _{-3.95\%}$ & 0.947 $ \pm 2.4\times10^{-3}$  $^{+2.2\%} _{-2.6\%}$    &  1.32  \\ \hline 
		
	ZH    & 0.608 $ \pm 5.9 \times10^{-4}$  $^{+3.11\%} _{-3.85\%}$ & 0.887$ \pm 1.4  \times10^{-3}$  $^{+2.0\%} _{-2.4\%}$ &  1.35   \\ \hline	
		
	\end{tabular} 
		\caption{The LO and NLO total cross section of VH production at  $\sqrt{s}= 14$ TeV. }
	\label{tab:1}
\end{table}

In table~\ref{tab:1}, we report our predictions results of the LO and NLO total cross sections and their corresponding K-factors for VH production. The impact of NLO QCD corrections  ranges from 32$\%$ to 38$\%$ of the LO term for all our processes, this is due to the real and virtual contributions introduced at NLO. We remind that the ZH production via the process gg $\rightarrow$ ZH contributes only at next-to-leading order (NLO) and its cross section is: $\sigma_{NLO}^{gg \rightarrow ZH}$ = 0.066$ \pm2.8\times10^{-4}$ $^{+24.3\%} _{-18.6\%}$ which represents 7$\%$ of the total cross section. The table also shows that the size of QCD corrections are large for process pp $\rightarrow$ W$^{-}$ H with the largest value of the factor K equal to 1.38.

Our total cross section predictions agree well with those presented in~\cite{ref:22}  at 13 TeV. The same is true when we compare them with experimental measurements given by the ATLAS Collaborations~\cite{ref:4} at $\sqrt{s}= 13$ TeV in pp collisions using data collected during 2015 and 2016: 

$\sigma_{WH}\cdot \mathcal{B}_{H \rightarrow WW^{\ast}}$= 0.67$^{+0.31}_{-0.27}$(stat.)$^{+0.14}_{-0.11}$(exp syst.)$^{+0.11}_{-0.09}$(theo syst.) pb,

$\sigma_{ZH} \cdot \mathcal{B}_{H \rightarrow WW^{\ast}}$= 0.54$^{+0.31}_{-0.24}$(stat.)$^{+0.10}_{-0.05}$(exp syst.)$^{+0.11}_{-0.05}$(theo syst.) pb.

with $\mathcal{B}_{H \rightarrow WW^{\ast}}$ is the branching-fraction values.

Further, we have simulated the scale uncertainty  by varying $\mu_{R}$ and $\mu_{F}$ simultaneously between 0.5$M_{H} $ < $\mu_{R}$, $\mu_{F}$ < 2$M_{H} $ with the constraint 0.5 < $\mu_{R}$/$\mu_{F}$ < 2. We note that the scale uncertainty at NLO is smaller than the corresponding uncertainty of the LO results.

\subsection{Distributions }
\label{subsec:3.2}

\begin{figure}
	\centering	
	\includegraphics[width=.30\textwidth,trim=0 0 0 0,clip]{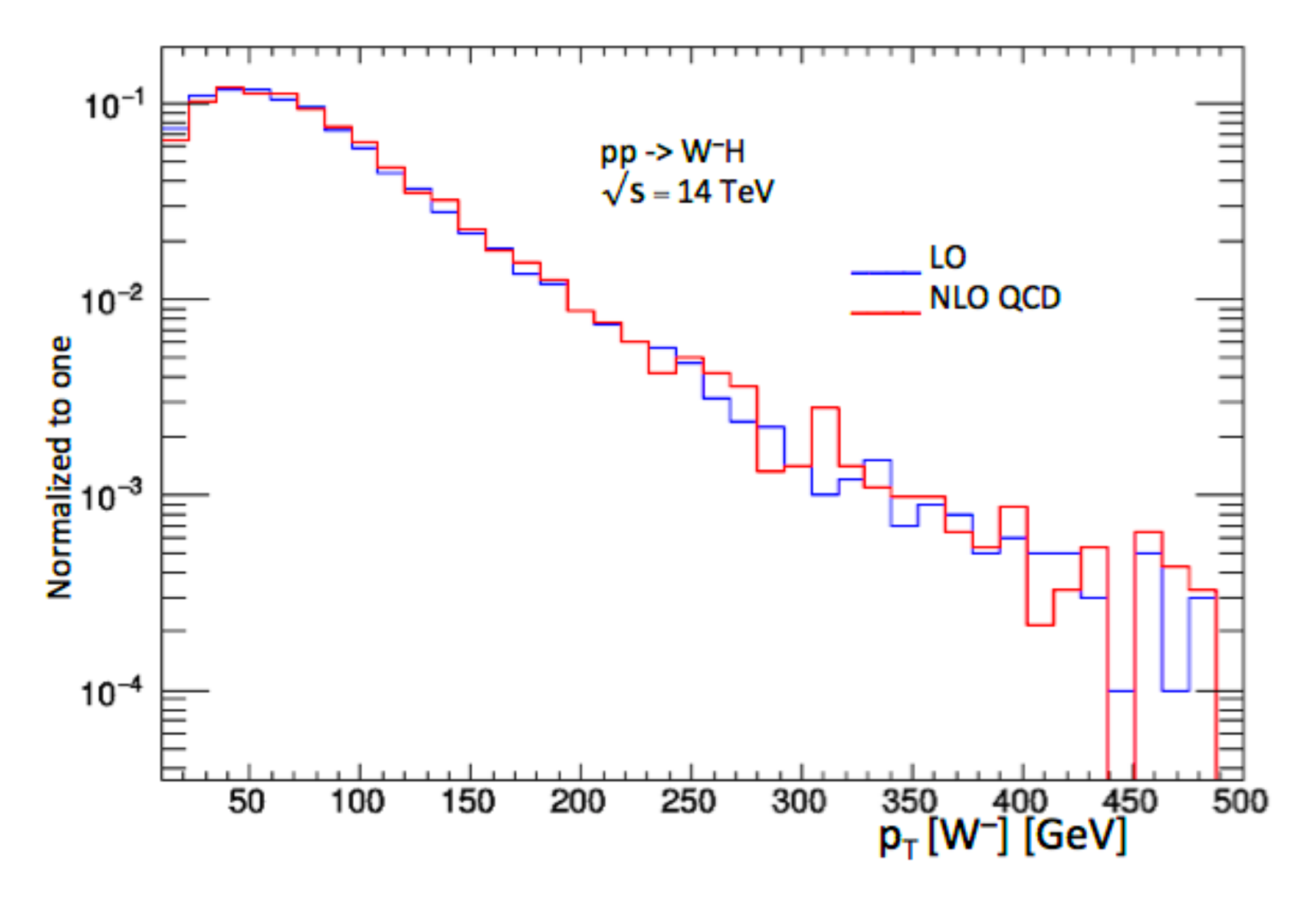}	\includegraphics[width=.30\textwidth,trim=0 0 0 0,clip]{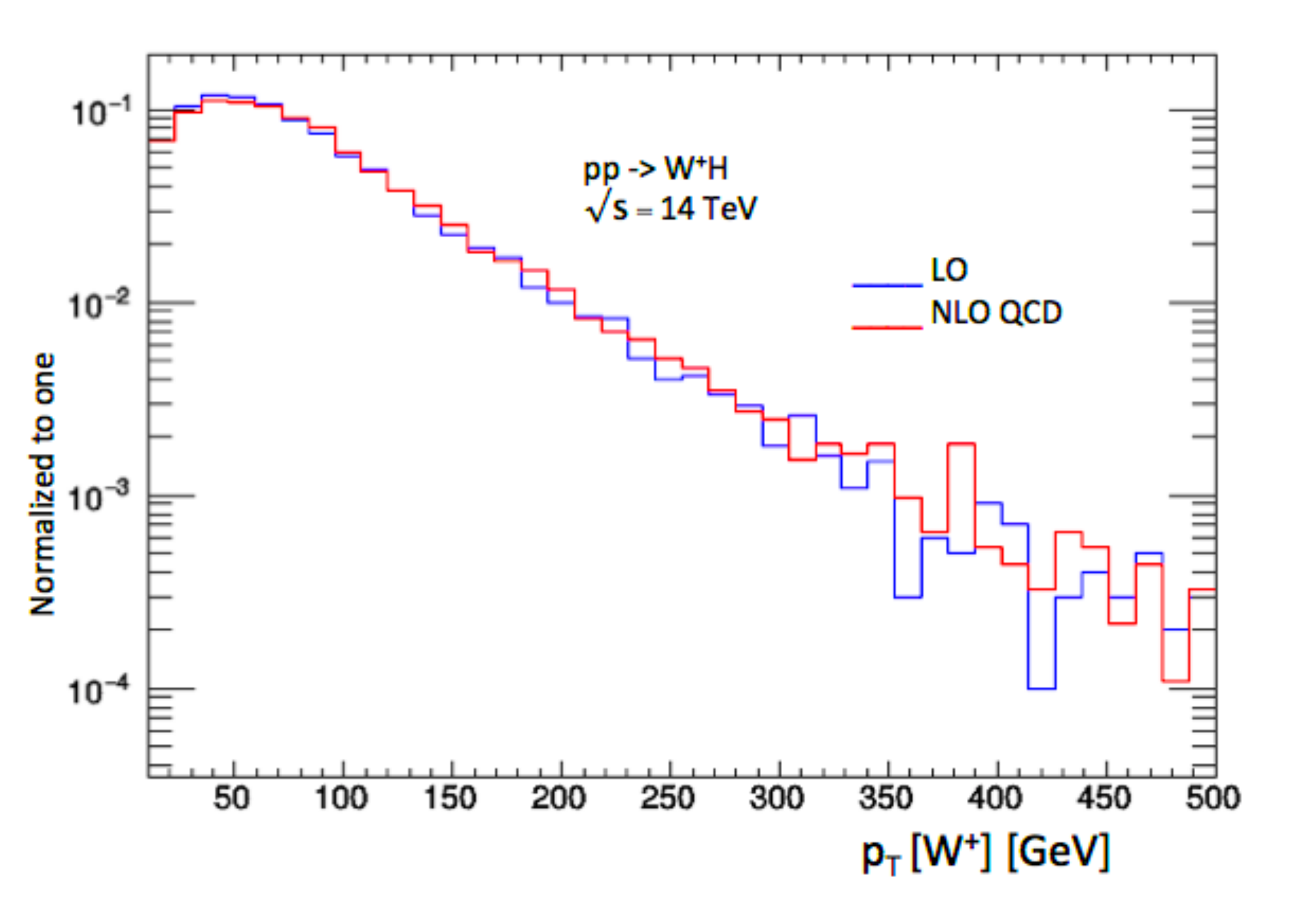} 	
	\includegraphics[width=.30\textwidth,trim=0 0 0 0,clip]{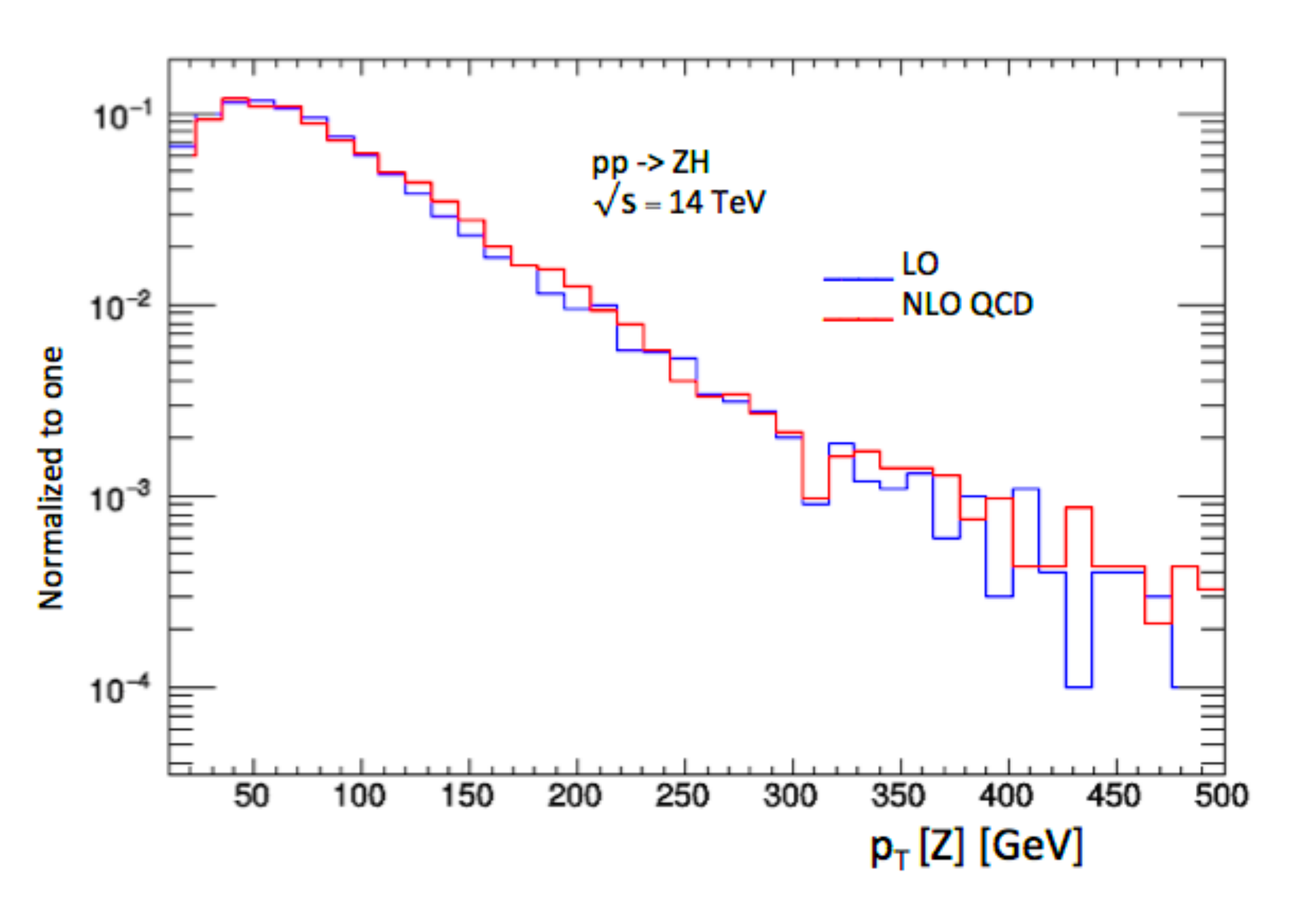} \\

	\caption{The LO and NLO transverse momentum distributions of vector bosons for VH productions at 14 TeV.
		\label{fig:4}
	}	
\end{figure}

Now, we investigate various kinematical distributions. we begin by the vector bosons transverse momentums at partonic level as plotted in figure~\ref{fig:4}, we see that the LO and NLO QCD corrections distributions have a same shape for all processes pp $\rightarrow$ VH (V= W$^{\pm}$ and Z ).  We estimate two regions, the first where the distributions grow lightly at low values of transverse momentums $p_{T} (V) <$ 60 GeV, then in the second, it decreases for the rest $p_{T} (V)$.

\begin{figure}
	\centering	
	\includegraphics[width=.30\textwidth,trim=0 0 0 0,clip]{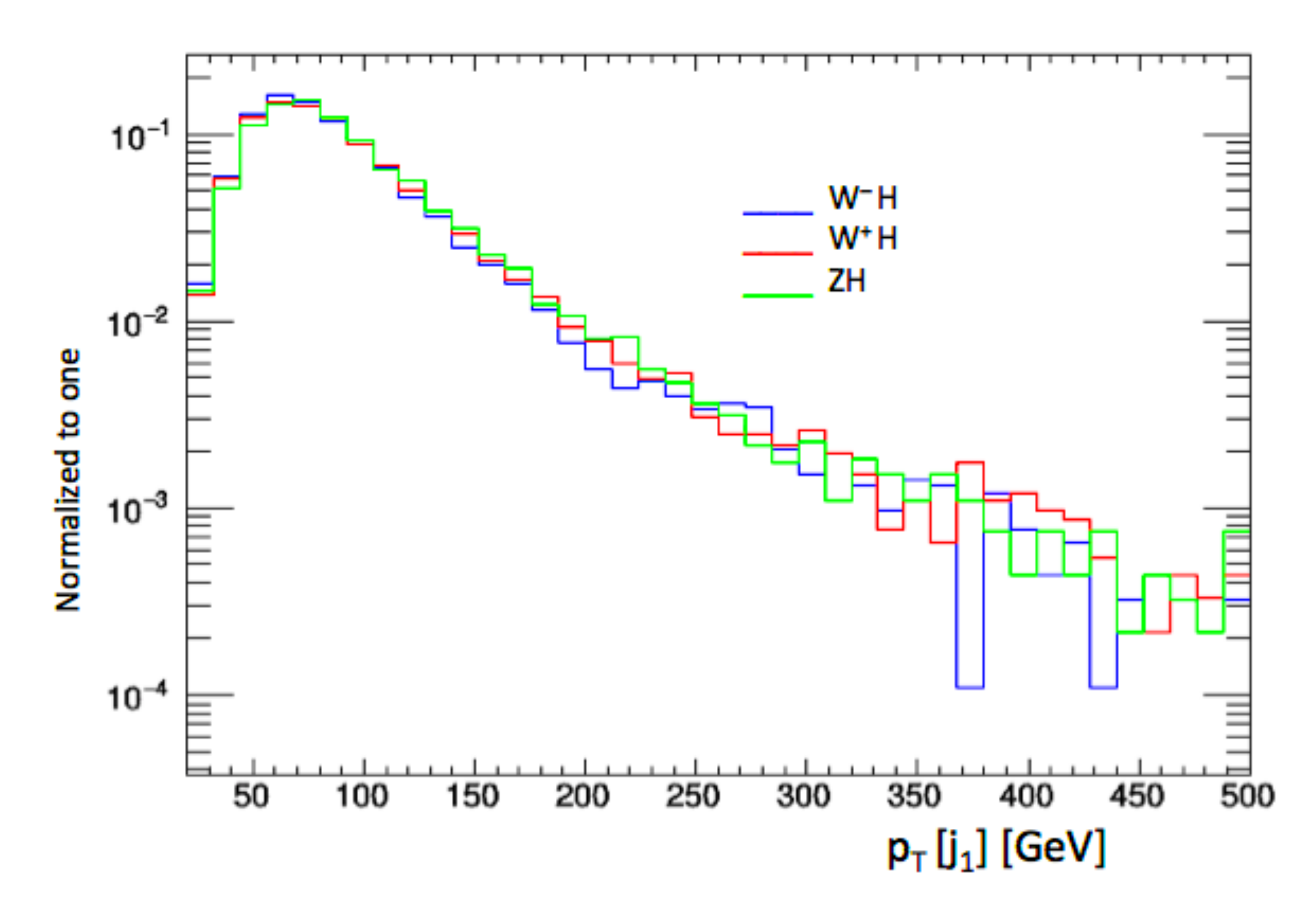}	\includegraphics[width=.30\textwidth,trim=0 0 0 0,clip]{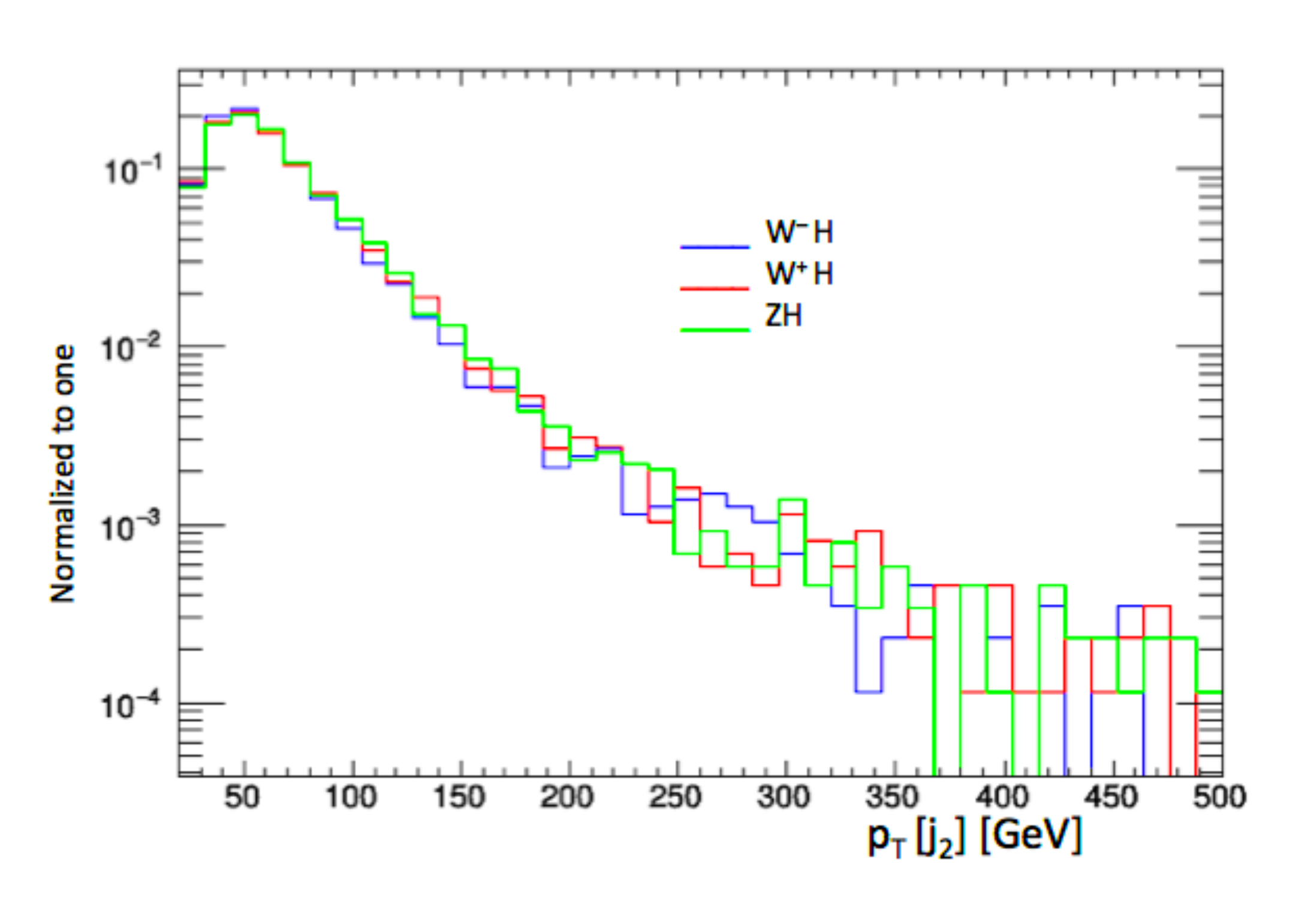} 	
	\includegraphics[width=.30\textwidth,trim=0 0 0 0,clip]{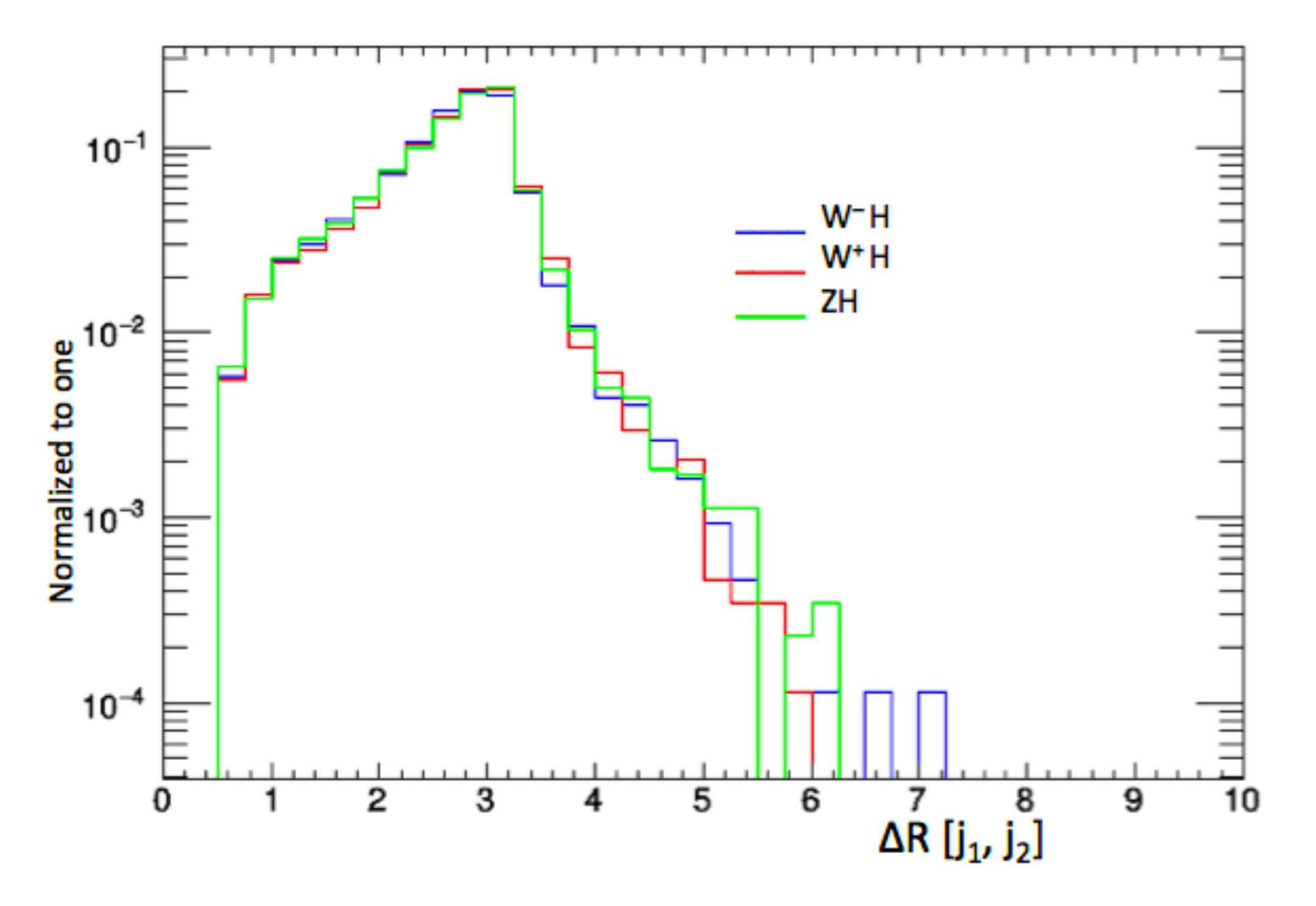} \\

	\caption{Transverse momentum distributions of the two tagging jets reconstructed at NLO-QCD and the corresponding separation $\Delta R$ for VH productionsat 14 TeV.
		\label{fig:5}	}	
\end{figure}

Transverse momentum distributions of the two reconstructed tagging jets at NLO-QCD  and $\Delta R$ distance (a separation between two hardest jets) are reported in figure~\ref{fig:5}.  The behaviour of the hardest jets distributions has overall the aspect similar for all processes, indeed at low $p_{T} (j_{1}) <$ 80 GeV and $p_{T} (j_{2}) <$ 50 GeV, the distributions are largest, then it increases slightly. The next-to-hardest jet distribution vanishes rapidly comparing to the leading jet. Additionally the NLO QCD put on the distributions a peak at $\Delta R$ around 3.

\begin{figure}
	\centering	
	\includegraphics[width=.30\textwidth,trim=0 0 0 0,clip]{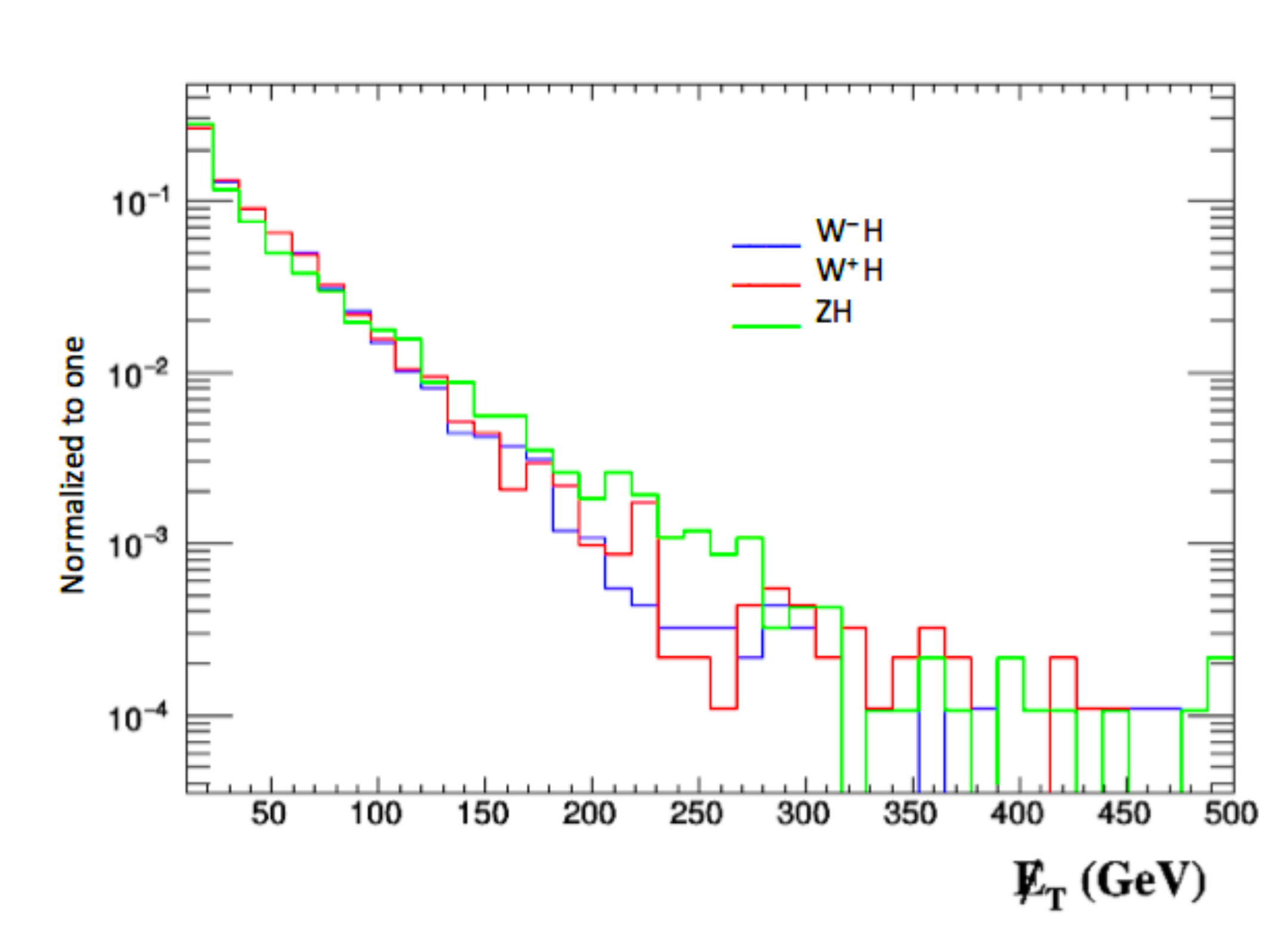}

	\caption{Rapidity distributions of Z-boson pair reconstructed after showering and hadronization with PYTHIA 8 for the processes~\ref{eq:1} at 14 TeV.
		\label{fig:6}
	}	
\end{figure}

The figure~\ref{fig:6} shows the E$_{T}^{miss}$ distribution which decreases with increment of E$_{T}^{miss}$. From E$_{T}^{miss} >$ 150 GeV, the signal for ZH production is greater than W$^{\pm}$H ones, this is due to the top-quark-loop included by the gg fusion.

\section{Conclusion }
\label{sec:4}

We have simulated the production of the Higgs boson associated with vector bosons W$^{\pm}$ and Z in proton-proton collisions at $\sqrt{s}$ = 14 TeV. We have illustrated the effects of NLO QCD corrections on this production. These laters produce new channels like gluon-gluon fusion which only appears at this level for ZH production. NLO QCD corrections also affect on the calculation of the total cross sections and we have found that these corrections are of the order of 38$\%$ of the first term. 
Scale uncertainties have equally been estimated, they vary from -4.37$\%$ to +3.58$\%$.

We have shown the shape of the spectrum for vector boson transverse momentums at parton level which are similar for all processes pp $\rightarrow$ VH (W$^{\pm}$, Z). The transverse momentum of the two tagging jets reconstructed at the first perturbative order allows us to confirm that the distribution is more sizeable at $p_{T} (j_{1})<$  M$_{V+H}$. For all processes VH production the maximum of the separation distance between two hardest jets is at $\Delta R$ = 3. We finally have pointed out the behaviour of missing energy transverse. 

We conclude that this study is an interesting analysis and we can expand it in different channels as H$\rightarrow b\bar{b}$ and exploit the information in new physics searches, We leave this to investigate in the near future.

\section*{Acknowledgements}

This work was realized with the support of the Algerian Ministry of Higher Education and Scientic Research.


\begin{thebibliography}{99}






\bibitem{ref:1}

The ATLAS Collaboration, \emph{Observation of a New Particle in the Search for the Standard Model Higgs Boson with the ATLAS Detector at the LHC}, \emph{Phys. Lett. B} (2012) arXiv:hep-ex/1207.7214.


\bibitem{ref:2}

The CMS Collaboration, \emph{Observation of a new boson at a mass of 125 GeV with the CMS experiment at the LHC}, \emph{Phys. Lett. B} (2013) arXiv:hep-ex/1207.7235.



\bibitem{ref:3}

The ATLAS Collaboration, \emph{Study of (W/Z)H production and Higgs boson couplings using $H \rightarrow WW^{\ast}$ decays with the ATLAS detector}, \emph{JHEP} (2015) arXiv:hep-ex/1506.06641.



\bibitem{ref:4}

The ATLAS Collaboration, \emph{Measurement of the production cross section for a
Higgs boson in association with a vector boson in the $H \rightarrow WW^{\ast} \rightarrow l\nu l\nu $ channel in p p collisions at $\sqrt{s}$ = 13 TeV with the ATLAS detector}, \emph{Phys. Lett. B} (2019) arXiv:hep-ex/1903.10052.



\bibitem{ref:5}

The ATLAS Collaboration, \emph{Measurements of WH and ZH production in the
$H \rightarrow b\bar{b}$ decay channel in p p collisions at 13TeV with the ATLAS detector}, \emph{EPJC} (2020) arXiv:hep-ex/2007.02873.



\bibitem{ref:6}

The CMS Collaboration, \emph{Combination of searches for heavy resonances decaying to WW, WZ, ZZ, WH, and ZH boson pairs in proton-proton collisions at $\sqrt{s}$ = 8 and 13 TeV}, \emph{Phys. Lett. B} (2017) arXiv:hep-ex/1705.09171.


\bibitem{ref:7}

C. Zecher, T. Matsuura and J.J. van der Bij, \emph{Leptonic Signals from off-shell Z Boson Pairs at Hadron Colliders}, \emph{Z.Phys. C} (1994) arXiv:hep-ph/9404295.


\bibitem{ref:8}

Q. Hong Cao, C. Sheng Li and S. Hua Zhu, \emph{Leading Electroweak Corrections to the Neutral Higgs Boson Production at the Fermilab Tevatron}, \emph{Commun. Theor. Phys} (2000) arXiv:hep-ph/9810458.


\bibitem{ref:9}

H. Baer, B. Bailey, and J. F. Owens, \emph{ $O(\alpha_{s})$ Monte Carlo approach to W + Higgs associated production at hadron supercolliders}, \emph{Phys. Rev. D} (1993).


\bibitem{ref:10}

A. Banfi and J. Cancinob, \emph{Implications of QCD radiative corrections on high-pT Higgs searches}, \emph{Phys. Lett. B} (2012) arXiv:hep-ph/1207.0674.



\bibitem{ref:11}

M. L. Ciccolini, S. Dittmaier and M. Kramer, \emph{Electroweak Radiative Corrections to Associated WH and ZH Production at Hadron Colliders}, \emph{Phys.Rev. D} (2003) arXiv:hep-ph/0306234.


\bibitem{ref:12}

O. Brein, A. Djouadi and R. Harlander, \emph{NNLO QCD corrections to the Higgs-strahlung processes at hadron colliders}, \emph{Phys. Lett. B} (2004) arXiv:hep-ph/0307206.




\bibitem{ref:13}

G. Ferrera, M. Grazzini and F. Tramontano, \emph{Higher-order QCD effects for associated WH production and decay at the LHC}, \emph{JHEP} (2014) arXiv:hep-ph/1312.1669.




\bibitem{ref:14}

R. Gauld, A. Gehrmann-De Ridder, E. W. N. Glover, A. Huss and I. Majer, \emph{Precise predictions for WH+jet production at the LHC} (2020) arXiv:hep-ph/2009.14209.




\bibitem{ref:15}

G. Ferrera, M. Grazzini and F. Tramontano, \emph{Associated ZH production at hadron colliders: the fully differential NNLO QCD calculation}, \emph{Phys. Lett. B} (2014) arXiv:hep-ph/1407.4747.




\bibitem{ref:16}

F. M. Renard, \emph{Test of Higgs boson compositeness in ZH production through gluon-gluon and photon-photon collisions} (2017) arXiv:hep-ph/1701.09116.





\bibitem{ref:17}

J. Alwall, M. Herquet, F. Maltoni,O. Mattelaerc and T. Stelzerd, \emph{MadGraph 5 : Going Beyond}, \emph{JHEP}(2011) arXiv:hep-ph/1106.0522.

\bibitem{ref:18}
T. Sjostrand, S. Mrenna, and P. Z. Skands, \emph{A Brief Introduction to PYTHIA 8.1}, \emph{Comput.Phys.Commun} ( 2007) arXiv:hep-ph/0710.3820.


\bibitem{ref:19}

G.'t Hooft and M. Veltman, \emph{Regularization and Renormalization of Gauge Fields}, \emph{Nucl.Phys. B}(1972)


\bibitem{ref:20}
The NNPDF Collaboration: R. D. Ball, V. Bertone, S. Carrazza, C. S. Deans, L. D. Debbio, S. Forte, A. Gu anti, N. P. Hartland, J. Latorre, J. Rojo and M. Ubiali, \emph{Parton distributions with LHC data}, \emph{Nucl. Phys. B} (2012) arXiv: hep-ph/1207.1303.

\bibitem{ref:21}
M. Cacciari, G. P. Salam, and G. Soyez, \emph{The anti-kt jet clustering algorithm}, \emph{JHEP} (2008) arXiv:hep-ph/1111.6097. 

\bibitem{ref:22}
J. Alwall, R. Frederix, S. Frixione, V. Hirschi, F. Maltoni, O. Mattelaer, H.-S. Shao, T. Stelzer, P. Torrielli and M. Zaro, \emph{The automated computation of tree-level and next-to-leading order differential cross sections, and their matching to parton shower simulations}, \emph{JHEP} (2014) arXiv:hep-ph/1405.0301. 







































































































\end{thebibliography}


\end{document}